\documentclass[epj,final]{svjour}

\usepackage[dvips]{graphicx}
\usepackage{hyperref}

\begin{document}

\title{Angular analysis of bremsstrahlung in $\alpha$-decay}
% \subtitle{Angular bremsstrahlung in $\alpha$-decay}
%
\dedication{This paper is dedicated to the memory of
Dr.~Ivan~Egorovich~Kashuba ---
a brilliant scientist with bright nature
who worked in science till his last days.
}

\author{Sergei~P.~Maydanyuk\thanks{maidan@kinr.kiev.ua}%
\and
Vladislav~S.~Olkhovsky\thanks{olkhovsk@kinr.kiev.ua}}
%
% \offprints{Sergei~P.~Maydanyuk}
\mail{Sergei~P.~Maydanyuk}
\institute{Institute for Nuclear Research, %
National Academy of Sciences of Ukraine\\%
prosp. Nauki, 47, Kiev-28, 03680, Ukraine}
\date{Received: 9 March 2006 \newline
Published online: 14 July 2006 ---
Societa Italiana di Fisica / Springer-Verlag 2006 \newline
Communicated by G.~Orlandini}
% \date{\today}
%---------------------------------------------------------------------------

%---------------------------------------------------------------------------
\abstract{
\rm
A new quantum electrodynamical method of calculations of bremsstrahlung
spectra in the $\alpha$-decay of heavy nuclei taking into account the
angle between the directions of $\alpha$-particle motion (or its
tunneling) and photon emission is presented.
The angular bremsstrahlung spectra for $^{210}\mbox{Po}$ have
been obtained for the first time.
According to calculations, the bremsstrahlung in the $\alpha$-decay
of this nucleus depends extremely weakly on the angle. Taking into account
nuclear forces, such dependence is not changed visibly.
An analytical formula of the angular dependence of the bremsstrahlung
spectra is proposed and gives its harmonic behavior.
The extremal values of the angle, at which the bremsstrahlung has
maximal and minimal values, has been found.
% An accuracy of the method is maximal inside the low-energy photon
% region.
%
}
\PACS{%
{23.60.+e}{Alpha decay} \and
{41.60.-m}{Radiation by moving charges} \and
{23.20.Js}{Multipole matrix elements
           (in electromagnetic transitions)} \and
{03.65.Xp}{Tunneling, traversal time, quantum Zeno dynamics}
%{27.80.+w}{A is greater than or equal to 190 and is less than or equal 
%to 219 (properties of specific nuclei listed by mass ranges)}
}
%---------------------------------------------------------------------------

%---------------------------------------------------------------------------
% \arxiv{nucl-th/0408022}

\maketitle
\hugehead

% \newpage
% \tableofcontents
% ***************************************************************************

% ***************************************************************************
\section{Introduction
\label{introduction}}

Experiments
\cite{D'Arrigo.1994.PHLTA,Kasagi.1997.JPHGB,Kasagi.1997.PRLTA}
with measurements of bremsstrahlung (Br) spectra in the $\alpha$-decay
of the nuclei $^{210}\mbox{Po}$, $^{214}\mbox{Po}$, $^{226}\mbox{Ra}$ and
$^{244}\mbox{Cm}$ have caused an increased interest. One of the key
ideas of the fulfillment of such experiments consists in finding a method
of extraction of a new information about $\alpha$-decay dynamics 
from the Br spectra (and a detailed information about the dynamics of
tunneling).
One can note a certain difference between the Br spectra
\cite{D'Arrigo.1994.PHLTA} and
\cite{Kasagi.1997.JPHGB,Kasagi.1997.PRLTA} for $^{210}\mbox{Po}$,
obtained experimentally for the values $90^{\circ}$ and $25^{\circ}$ of the
angle between the directions of the $\alpha$-particle propagation and the
photon emission (these experiments and the difference of their spectra
are discussed in~\cite{Eremin.2000.PRLTA,Kasagi.2000.PRLTA}).
One can explain such difference between the Br angular spectra on the
basis of the following idea: \emph{the Br intensity depends on the
directions of emission of the photons and motion (with possible tunneling)
of the $\alpha$-particle relatively the daughter nucleus}.
In such a way, a three-dimensional picture of the $\alpha$-decay with
the accompanying Br in the spatial region of nuclear boundaries has
been devised.

However, if the Br intensity varies enough visibly with changing the
angle value, then one can suppose, that the photon emission is able
to influence essentially on the $\alpha$-decay dynamics and,
therefore, to change all of its characteristic.
From this point of view, the discussions
\cite{Eremin.2000.PRLTA,Kasagi.2000.PRLTA} open a way for obtaining
a new information about the $\alpha$-decay --- through the \emph{angular
analysis of the Br during the $\alpha$-decay}.
But for such researches a model describing the Br in the
$\alpha$-decay, which takes into account the value of the angle between
the directions of the $\alpha$-particle propagation (or tunneling) and
the photon emission, is needed.

In theoretical aspect, some progress has been made here.
One can note models of calculations of the Br spectra in the $\alpha$-decay,
developed on the basis of quantum electrodynamics with use of
perturbation theory:
the first paper \cite{Batkin.1986.SJNCA} where
a general quantum-mechanical formalism of the calculation of the Br
spectra in the $\alpha$-decay is proposed and the Br spectrum for
$^{210}\mbox{Po}$ inside the photons energy region up to 200 keV was
estimated (even until the fulfillment of the first experiments);
essentially improved models in the dipole approximation
\cite{Papenbrock.1998.PRLTA,Takigawa.1999.PHRVA} and in the
multipolar expansion \cite{Tkalya.1999.PHRVA} of photons current
(wave function) with application of the \emph{Fermi golden rule};
an approach \cite{Kurgalin.2004} of the calculation of the Br spectra
with realistic barriers of the $\alpha$-decay),
models \cite{Dyakonov.1996.PRLTA,Dyakonov.1999.PHRVA,%
Bertulani.1999.PHRVA,Takigawa.1999.PHRVA} developed in semiclassical
approximation (see also the Br spectra calculations
in~\cite{Kasagi.1997.PRLTA}),
\emph{instant accelerated models}
\cite{D'Arrigo.1994.PHLTA,Tkalya.1999.PHRVA} constructed on the
basis of classical electrodynamics
(see also \cite{Dyakonov.1999.PHRVA}),
methods \cite{Bertulani.1999.PHRVA,Misicu.2001.JPHGB,Dijk.2003.FBSSE,%
Dyakonov.1996.PRLTA,Dyakonov.1999.PHRVA}, directed on a nonstationary
description of the $\alpha$-decay with the accompanying Br and the
calculations of such non-stationary characteristics as \emph{tunneling
time}. One can recall also papers
\cite{Serot.1994.NUPHA,Dijk.1999.PRLTA,Dijk.2002.PHRVA,Ivlev.2004.PHRVA}
with study of dynamics of subbarrier tunneling in the $\alpha$-decay;
an effect, opened in \cite{Flambaum.1999.PRLTA} and named 
\emph{M\"{u}nchhausen effect}, which increases the barrier
penetrability due to charged-particle emission during its tunneling
and which can be extremely interesting for further study of the photon
bremsstrahlung during subbarier tunneling in the $\alpha$-decay).
However, one needs to say that at this stage the calculations 
of the Br spectra by all these approaches are reduced to 
obtaining their integral (or averaged by angles) values and, 
therefore, they do not allow to fulfill an angular analysis of the
experimental Br spectra (here, one can quote an
approach in~\cite{Misicu.2001.JPHGB} based on classical
electrodynamics, which shows a way for obtaining the angular spectra
(see (15) and (17), p.~999), however here we shall use the direct
quantum-mechanical approach of the Br spectra calculation, which
describes the quantum effect of the \emph{subbarrier
Br} more precisely).

In~\cite{Maydanyuk.2003.PTP} we had developed a multipolar method
which takes into account the angle between the directions of the
$\alpha$-particle propagation and the photon emission. But the angular
integrals used in this method are difficult enough to be obtained and
some approximations are used, there was a convergence problem with the
calculations of multipoles of larger orders, only the angular dependences
of matrix elements of the selected multipoles E1 and M1 were found,
while it is interesting to know also the angular dependence of
the total Br spectra.
Moreover, computer calculations of the Br spectra and their angular
analysis will be essentially more complicated, if one passes from the
type of the potential used in this paper (and also
in~\cite{Papenbrock.1998.PRLTA,Tkalya.1999.PHRVA}) to
realistic potentials. In this sense, the approach proposed
in~\cite{Maydanyuk.2003.PTP} is not convenient enough.

In this paper we present a new method for the angular calculations of
the Br spectra in the $\alpha$-decay of nuclei (started in 
\cite{Maydanyuk.nucl-th.0404013}, with the resolution of a convergence
problem in the Br spectra calculations existing in
\cite{Maydanyuk.nucl-th.0404013}).
In our approach we introduce a simplified transformation, which reduce 
the complicated angular formalism of the calculation of the Br spectra
(presented in~\cite{Maydanyuk.2003.PTP}) to a maximally simple form
(this makes the method clearer and more \underline{comprehensible}), 
with keeping the calculating accuracy as good as possible, where
only one angle from all angular parameters is used --- the angle
used in experiments
\cite{Eremin.2000.PRLTA,Kasagi.2000.PRLTA} for $^{210}\mbox{Po}$.
We show, that this proposed transformation works in the low-energy region
of photons and, therefore, it can be applied to the analysis of all up to
now existing experimental data of Br spectra in the $\alpha$-decay of
spherical nuclei (we relate $^{210}\mbox{Po}$ to them).
In this paper we present the results of calculations of the angular
spectra of the Br in the $\alpha$-decay for $^{210}\mbox{Po}$
(the Br angular spectra in the $\alpha$-decay have been obtained for the
first time).
We show, how the Br spectra are changed after deformation of the form
of the $\alpha$-decay barrier as a result of the correction of a
component of the $\alpha$-nucleus potential of nuclear forces 
(we did not find such calculations in other papers).
An analysis of the problem of the calculations convergence of the Br
spectra in the $\alpha$-decay is included in the paper, whose resolution
plays a key role for obtaining of the reliable values of the spectra.
% ***************************************************************************

% ***************************************************************************
\section{A formalism of the calculations of the bremsstrahlung
spectra in the stationary approach
\label{sec.2}}

We shall consider the decay of the nucleus as the decay of the
compound quantum system: $\alpha$-particle and daughter nucleus.
The $\alpha$-particle is the electrically charged particle and
during its motion inside the electromagnetic field of the daughter
nucleus it emits photons. The spontaneous emission of the photon
changes a state of the compound system, which is described by its
wave function.
For a quantitative estimation of the photon Br we use a transition
of the system from its state before the photon emission (we name such
state as the \emph{initial $i$-state}) into its state after the
photon emission (we name such state as the \emph{final $f$-state}).
One can define a matrix element of such transition of the system and
on its basis find the Br probability during the $\alpha$-decay (for
convenience, we denote it as $W(w)$). According to
\cite{Maydanyuk.2003.PTP}, we obtain:
\begin{equation}
\begin{array}{cccc}
%  \displaystyle\frac{dW(w)}{d\Omega_{\nu}} =
  W(w) =
    N_{0} k_{f} w \bigl| p(w) \bigr|^{2}, &
  N_{0} = \displaystyle\frac{Z_{\rm eff}^{2} e^{2}}{(2\pi)^{4} m}, &
  k_{i,f} = \sqrt{2mE_{i,f}}, &
  w = E_{i} - E_{f},
\end{array}
\label{eq.2.1}
\end{equation}
where $p(w)$ has a form 
\begin{equation}
  p(w) =
    \sum\limits_{\alpha = 1, 2}
    \mathbf{ e}^{(\alpha)*}
    \int\limits^{+\infty}_{0} dr \int
    r^{2} \psi^{*}_{f}(\mathbf{ r})
    e^{-i\mathbf{ kr}}
    \displaystyle\frac{\partial}{\partial \mathbf{ r}}
    \psi_{i}(\mathbf{ r}) d\Omega.
\label{eq.2.2}
\end{equation}
Here
$Z_{\rm eff}$ and $m$ are effective charge and reduced mass of the
system,
$E_{i,f}$, $k_{i,f}$ and $\psi_{i,f}(\mathbf{r})$ are total energy,
wave vector and wave function of the system in the initial $i$-state
or in the final $f$-state (in dependence on the index $i$ or $f$ in
use),
$\psi_{i}(\mathbf{r})$ and $\psi_{f}(\mathbf{r})$ are the wave
function of the system in the initial $i$- and the final $f$-states,
$\mathbf{e}^{(\alpha)}$ is polarization vector of the photon emitted,
$\mathbf{k}$ is photon wave vector,
$w = k = \bigl|\mathbf{k}\bigr|$ is photon frequency (energy).
The vector $\mathbf{e}^{(\alpha)}$ is perpendicular to $\mathbf{k}$
in Coulomb calibration.
We use such system of units: $\hbar = 1$ and $c = 1$. Notations are
used in accordance with \cite{Maydanyuk.2003.PTP}.
Similar expressions for the Br probability are used in 
\cite{Papenbrock.1998.PRLTA,Takigawa.1999.PHRVA,Tkalya.1999.PHRVA}
with further application of Fermi golden rule.

In accordance with main statements of quantum mechanics, the wave
functions of the system in the states before and after the photon
emission are defined inside all space region, including the region
of the subbarrier tunneling. A definition of a matrix element of
the transition of this system requires an account of all space
region of the definition of the wave functions of this system in two
states. Therefore, \emph{we should include the tunneling region into
the definition of the matrix element of the Br, irrespective of we
know, whether the photons emission is possible during tunneling or
not}.

Let's consider a subintegral expression in (\ref{eq.2.2}). Here, the
wave function $\psi_{i}(\mathbf{r})$ for the initial $i$-state and
the wave function $\psi_{f}(\mathbf{r})$ for the $f$-final state take
into account the directions of propagation (or tunneling) of the
$\alpha$-particle before the photon emission and after it,
correspondingly; the photons wave function (its main part consists in
the exponent $\exp{(-i\mathbf{ kr})}$) points to the direction of
propagation of the photon emitted.
We see, that the quantum mechanical approach for calculation of the
Br spectra initially has a detailed angular information about the
process of $\alpha$-decay with the accompanying Br.

However, we see that a further development of the approach for the
calculations of the Br spectra in the $\alpha$-decay on the basis of
the formulas (\ref{eq.2.1}) and (\ref{eq.2.2}) by other authors
(which consist in the calculations of $p(w)$) gives rise to angular
averaging of the spectra.
And a necesity has been arising in construction of an approach,
which allows simply enough to calculate the Br spectra
(with possible resolving the convergence problem in the computer
calculations)
with taking into account of the angle between the directions of the
$\alpha$-particle propagation (with possible tunneling) and the
photon emission and without (essential) decreasing of the accuracy.
% ***************************************************************************

% ***************************************************************************
\section{A simplified angular method of the calculations of the
matrix element
\label{sec.3}}

In \cite{Maydanyuk.nucl-th.0404013} the approach for the calculation
of the Br spectra, allowing to find a dependence of the total Br
spectra on the angle between the directions of the $\alpha$-particle
propagation (or tunneling) and the photon emission, was proposed.
However, further research has shown, that it is extremely difficult
to achieve a convergence in the computer calculations of the Br
spectra by such approach and, therefore, such a method requires an
essential development. Here, we propose a consecutive statement of
such approach with a resolution of the convergence problem.

Let's rewrite vectors $\mathbf{e}^{\alpha}$ of polarization through
the vectors $\mathbf{\xi}_{-1}$ and $\mathbf{\xi}_{+1}$ of circular
polarization with opposite directions of rotation
(see~\cite{Eisenberg.1973}, p.~42):
\begin{equation}
\begin{array}{ll}
  \mathbf{\xi}_{-1} = \displaystyle\frac{1}{\sqrt{2}}
                   (\mathbf{e}^{1} - i\mathbf{e}^{2}), &
  \mathbf{\xi}_{+1} = -\displaystyle\frac{1}{\sqrt{2}}
                   (\mathbf{e}^{1} + i\mathbf{e}^{2}).
\end{array}
\label{eq.3.1}
\end{equation}
Substituting these values into (\ref{eq.2.2}), we obtain:
\begin{equation}
  p(w) =
        \sum\limits_{\mu = -1, 1}
        h_{\mu}\mathbf{\xi}^{*}_{\mu}
        \int\limits^{+\infty}_{0} dr \int
        r^{2} \psi^{*}_{f}(\mathbf{r})
        e^{-i\mathbf{kr}}
        \displaystyle\frac{\partial}{\partial \mathbf{r}}
        \psi_{i}(\mathbf{r}) d\Omega,
\label{eq.3.2}
\end{equation}
where
\begin{equation}
\begin{array}{lcr}

  h_{-1} = \displaystyle\frac{1}{\sqrt{2}} (1-i), &
  h_{1}  = - \displaystyle\frac{1}{\sqrt{2}} (1+i), &
  h_{-1} + h_{1} = -i \sqrt{2}.
\end{array}
\label{eq.3.3}
\end{equation}

Using the following properties (see~\cite{Eisenberg.1973} p.~44--46,
\cite{Maydanyuk.2003.PTP}):
\begin{equation}
\begin{array}{l}
  \displaystyle\frac{\partial}{\partial \mathbf{r}}
  \psi_{i}(\mathbf{r}) =
        -\displaystyle\frac{d\psi_{i}(r)}{dr}
        \mathbf{ T}_{01,0}(\mathbf{n}^{i}_{r}),       \\
  \mathbf{T}_{01,0}(\mathbf{n}^{i}_{r}) =
        \displaystyle\sum\limits_{\mu = -1,1} (110|-\mu\mu 0)
        Y_{1,-\mu}(\mathbf{n}^{i}_{r}) \mathbf{\xi}_{\mu}, \\
  (110|1, -1, 0) = (110|-1, 1, 0) = \sqrt{\displaystyle\frac{1}{3}},
\end{array}
\label{eq.3.4}
\end{equation}
where
$(110 | -\mu\mu 0)$ are Clebsch-Gordan coefficients and
$\mathbf{T}_{ll',\mu}(\mathbf{n})$ are vector spherical harmonics
(see~\cite{Eisenberg.1973}, p.~45 and we use quantum numbers $l=m=0$
in the initial $i$-state),
$Y_{1,\mu}(\mathbf{n}^{i,f}_{r})$ are normalized spherical functions
(see~\cite{Landau.1989}, p.~118--121 (28.7), p.~752--755),
we obtain:
\begin{equation}
  \displaystyle\frac{\partial}{\partial \mathbf{r}} \psi_{i}(\mathbf{r}) =
    -\displaystyle\frac{d\psi_{i}(r)}{dr}
    \sqrt{\displaystyle\frac{1}{3}}
    \sum\limits_{\mu^{\prime} = -1,1}
    Y_{1,-\mu^{\prime}}(\mathbf{n}^{i}_{r}) \mathbf{\xi}_{\mu^{\prime}}.
\label{eq.3.5}
\end{equation}
Taking into account (\ref{eq.3.2}), (\ref{eq.3.5}) and the
ortogonality condition of the vectors $\mathbf{\xi}^{*}_{\pm 1}$ and
$\mathbf{\xi}_{\mp 1}$ of polarization, we find:
\begin{equation}
  p(w) = -
        \sqrt{\displaystyle\frac{1}{3}}
        \sum\limits_{\mu = -1, 1}
        h_{\mu}
        \int\limits^{+\infty}_{0} dr
        r^{2} \psi^{*}_{f}(r)
        \displaystyle\frac{\partial \psi_{i}(r)}{\partial r}
        \int
        Y_{l'm'}^{*}(\mathbf{n}^{f}_{r})
        Y_{1,-\mu}(\mathbf{n}^{i}_{r})
        e^{-i\mathbf{k}\mathbf{r}}
        d\Omega,
\label{eq.3.6}
\end{equation}
where
$\psi_{f}(\mathbf{r}) =
  \psi_{f}(r) Y_{l^{\prime},m^{\prime}}(\mathbf{n}^{f}_{r})$.

Let's consider the vectors $\bf k$ and $\bf r$.
The vector $\bf k$ is an impulse of the photon, pointed out the
direction of its propagation.
The vector $\bf r$ is a radius-vector, pointed out a position of
the $\alpha$-particle relatively a center of mass of the daughter
nucleus and (because of mass of the daughter nucleus is larger
sufficiently than mass of the $\alpha$-particle) pointed out the
direction of its motion (or tunneling).
Then an angle between the vectors $\bf k$ and $\bf r$ (let's denote
it as $\beta$) is the angle between the direction
$\mathbf{n}_{r} = \mathbf{r}/r$ of motion (or tunneling) of 
the $\alpha$-particle and the direction
$\mathbf{n}_{ph} = \mathbf{k}/k$ of a propagation of the photon
emited, i.~e. it is the angle used in the experiments
\cite{Eremin.2000.PRLTA,Kasagi.1997.PRLTA,D'Arrigo.1994.PHLTA}.
% (with taking into account of a transition of expressions into системы центра 
% масс в лабораторную систему).
%
One can write
\begin{equation}
\begin{array}{lcr}
  \exp{(-i\mathbf{kr})} = \exp{(-ikr \cos{\beta})}, &
  k = |\mathbf{k}|, &
  r = |\mathbf{r}|.
\end{array}
\label{eq.3.7}
\end{equation}

Now we make such \underline{assumptions}:

\begin{itemize}
\vspace{-2mm}
\item
the photon emission process does not change the direction of motion
(or tunneling) of the $\alpha$-particle:
\begin{equation}
   \mathbf{n}^{i}_{r} = \mathbf{n}^{f}_{r},
\label{eq.3.8}
\end{equation}

\vspace{-2mm}
\item
the angle $\beta$ is not depended on the direction of the outgoing
$\alpha$-particle motion from the nucleus region.
\end{itemize}
Then, taking into account these assumptions and the ortogonality
property of the functions $Y_{lm}(\mathbf{n}_{r})$, we obtain the
following expression for $p(w,\beta) $:
\begin{equation}
  p(w, \beta) =
        -\sqrt{\displaystyle\frac{1}{3}}
        \sum\limits_{\mu = -1, 1}
        h_{\mu}
        \int\limits^{+\infty}_{0}
        r^{2} \psi^{*}_{f}(r)
        \displaystyle\frac{\partial \psi_{i}(r)}
        {\partial r}
        e^{-ikr \cos{\beta}}
        dr
\label{eq.3.9}
\end{equation}
and selection rules for quantum numbers $l$ and $m$ of the final
$f$-state:
\begin{equation}
\begin{array}{lll}
  \mbox{the initial state:} & l_{i} = 0, & m_{i} = 0;\\
  \mbox{the final state: } & l_{f} = 1, & m_{f} = -\mu = \pm 1.
\end{array}
\label{eq.3.10}
\end{equation}
% ***************************************************************************

% ***************************************************************************
\section{Spherical wave expansion
\label{sec.4}}

For further computer calculations of the integral (\ref{eq.3.9}),
let's use an expansion of the plane wave in the spherical waves
(for example, see \cite{Landau.1989}, p.~144, (34.1)):
\begin{equation}
  e^{ikz} =
    \displaystyle\sum\limits_{l=0}^{+\infty}
    (-i)^{l} (2l+1) P_{l} (\cos{\beta})
    \biggl(\displaystyle\frac{r}{k} \biggr)^{l}
    \biggl(\displaystyle\frac{1}{r} \displaystyle\frac{d}{dr} \biggr)^{l}
    \displaystyle\frac{\sin {kr}}{kr},
\label{eq.4.1}
\end{equation}
where $z=r \cos{\beta}$.
Introducing spherical Bessel functions (see \cite{Landau.1989},
p.~139, (33.9), (33.10) and (33.11)):
\begin{equation}
  j_{l}(kr) =
    (-1)^{l} \biggl(\displaystyle\frac{r}{k} \biggr)^{l}
    \biggl(\displaystyle\frac{1}{r} \displaystyle\frac{d}{dr} \biggr)^{l}
    \displaystyle\frac{\sin {kr}}{kr},
\label{eq.4.2}
\end{equation}
we obtain:
\begin{equation}
  e^{-ikr \cos{\beta}} =
  \biggl(e^{ikr \cos{\beta}} \biggr)^{*} =
    \displaystyle\sum\limits_{l=0}^{+\infty}
    i^{l} (-1)^{l} (2l+1) P_{l} (\cos{\beta}) j_{l}(kr)
\label{eq.4.3}
\end{equation}
and from (\ref{eq.3.9}) we find:
\begin{equation}
  p(w, \beta) =
    -\sqrt{\displaystyle\frac{1}{3}}
    \displaystyle\sum\limits_{l=0}^{+\infty}
    i^{l} (-1)^{l} (2l+1) P_{l} (\cos{\beta})
    \sum\limits_{\mu = -1, 1} h_{\mu}
      J_{m_{f}}(l,w),
\label{eq.4.4}
\end{equation}
where
\begin{equation}
  J_{m_{f}}(l,w) =
    \int\limits^{+\infty}_{0}
    r^{2} \psi^{*}_{f}(r)
    \displaystyle\frac{\partial\psi_{i}(r)} {\partial r}
    j_{l} (kr) dr.
\label{eq.4.5}
\end{equation}
$J_{m_{f}}(l,w)$ is a radial integral, not depended on the angle
$\beta$.
Now \emph{we obtain an explicit analytical dependence of the matrix
element $p(w,\beta)$ on the angle $\beta$ between the directions of
the $\alpha$-particle propagation and the photon emission}
(for the first time, in \cite{Maydanyuk.2003.PTP} the angular integrals
were obtained for selected components of this matrix element ---
multipoles E1 and M1 only, with use of more difficult calculations).
% ***************************************************************************

% ***************************************************************************
\section{The bremsstrahlung in the Coulomb field
\label{sec.5}}

Practically, in the numerical calculation of the Br spectra it is
convenient to divide the whole region of the integration into two
parts: the region 1 of a joint action of the Coulomb and nuclear
forces not far from the nucleus and the region 2, in which one can
neglect by the action of the nuclear forces in comparison with the
action of the Coulomb forces.
Our analysis has shown, that an attainment of the \emph{convergence}
of the Br spectra calculations (which determines their \emph{accuracy},
\emph{reliability} of the found Br spectra) is reached first of all
by correctness of the calculations in the region 2.
Namely, in this region it needs to solve a problem with definition of
the external boundary of integration (its increasing leads to
increasing of the accuracy of the obtained spectra, but to increasing
of difficulty of the calculations and analysis),
to choose the most effective method of the numerical integration
(of an improper integral with an oscillating and weakly damping
sub-integral function),
to solve a problem with attainment of needed accuracy and convergence
of the calculations.
It defines time of the calculations, minimization of which appears
extremely important for fulfillment of the real analysis of the
obtained Br spectra in dependence on needed parameters.
Therefore, maximal simplification of the formulas for the Br spectra
in the region 2 is useful.

Let's assume, that the potential, used in the radial integral
(\ref{eq.4.5}), in spatial region of $r$ is \emph{Coulomb} since the
value $R_{c}$. We accept $R_{c}$ as the internal boundary of the
region 2.
One can write the radial integral $J(l,w)$ in (\ref{eq.4.5}) so:
\begin{equation}
  J_{m_{f}}(l,w) = J_{in, m_{f}}(l,w) + J_{c}(l,w),
\label{eq.5.1}
\end{equation}
where
\begin{equation}
\begin{array}{lcl}
  J_{in, m_{f}}(l,w) & = &
    \displaystyle\int\limits^{R_{c}}_{0}
    r^{2} \psi^{*}_{f}(r, m_{f})
    \displaystyle\frac{\partial\psi_{i}(r)} {\partial r}
    j_{l} (kr) dr, \\
  J_{c}(l,w) & = &
    \displaystyle\int\limits^{+\infty}_{R_{c}}
    r^{2} \psi^{*}_{f}(r)
    \displaystyle\frac{\partial\psi_{i}(r)} {\partial r}
    j_{l} (kr) dr.
\end{array}
\label{eq.5.2}
\end{equation}
The radial integral $J_{c}(l,w)$ does not depend on the quantum
number $m$ of the systems in the final $f$-state. Then, one can
write $p(w,\beta)$ so (with taking into account (\ref{eq.3.3}) for
the Coulomb component):
\begin{equation}
  p(w,\beta) = p_{in}(w,\beta) + p_{c}(w,\beta),
\label{eq.5.3}
\end{equation}
where
\begin{equation}
\begin{array}{lcl}
  p_{in}(w,\beta) & = &
    -\sqrt{\displaystyle\frac{1}{3}}
    \displaystyle\sum\limits_{l=0}^{+\infty}
      i^{l} (-1)^{l} (2l+1) P_{l} (\cos{\beta})
    \displaystyle\sum\limits_{\mu = -1, 1} h_{\mu}
      J_{in, m_{f}}(l,w), \\
  p_{c}(w,\beta) & = &
    \sqrt{\displaystyle\frac{2}{3}}
    \displaystyle\sum\limits_{l=0}^{+\infty}
    i^{l+1} (-1)^{l} (2l+1) P_{l} (\cos{\beta}) J_{c}(l,w).
\end{array}
\label{eq.5.4}
\end{equation}
We see, that there is no any interference between the components
$p_{in}(w,\beta)$ and $p_{c}(w,\beta)$ in the calculations of the
total value of $p(w,\beta)$, but it exists in calculations of the
total Br spectra.
% ***************************************************************************

% ***************************************************************************
\section{The first approximation at $l=0$
\label{sec.6}}

% Найдем нулевое приближение вероятности ТИ в кулоновском поле при
% $l=0$.

Legandre's polynomial of the order $l$ equals (for example, see 
\cite{Landau.1989}, p.~752 (c.1)):
\begin{equation}
\begin{array}{cccc}
  P_{l}(\theta) =
    \displaystyle\frac{1}{2^{l} l!}
      \displaystyle\frac{d^{l}}{(d\theta)^{l}}
      (\theta^{2} - 1)^{l}, &
  P_{0}(\theta) = 1, &
  P_{1}(\theta) = \theta, &
  \theta = \cos{\beta}.
\end{array}
\label{eq.6.1}
\end{equation}
Then at $l=0$ we find:
\begin{equation}
\begin{array}{lcl}
  p_{in}^{(l=0)} (w,\beta) & = &
    -\sqrt{\displaystyle\frac{1}{3}}
    \displaystyle\sum\limits_{\mu = -1, 1} h_{\mu}
      J_{in, m_{f}}(0,w), \\
  p_{c}^{(l=0)} (w,\beta) & = &
    i \sqrt{\displaystyle\frac{2}{3}} J_{c}(0,w).
\end{array}
\label{eq.6.2}
\end{equation}

If for nuclei $^{210}\mbox{Po}$, $^{214}\mbox{Po}$, $^{226}\mbox{Ra}$ to use the
potential with parameters as in~\cite{Maydanyuk.2003.PTP} (and as
in~\cite{Papenbrock.1998.PRLTA,Tkalya.1999.PHRVA} also), then we
find, that Br from the internal spatial region till to $R_{c}$ is
extremely small ($p_{in}(w,\beta) << p_{c}(w,\beta)$). According to
our estimations, for such potential it is smaller in
$10^{-22}$--$10^{-24}$ times then Br from the external region.
Therefore, one can neglect by Br from the internal region, and the
total Br can be determined by the Coulomb field inside the barrier
region and the external region.
From (\ref{eq.2.1}) we write down the Br probability in the first
approximation at $l=0$:
\begin{equation}
  W_{l=0}(w) =
    N_{0} k_{f} w \bigl| p_{c}^{(l=0)}(w,\beta) \bigr|^{2} =
    \displaystyle\frac{2}{3} N_{0} k_{f} w \bigl| J_{c}(0,w) \bigr|^{2}.
\label{eq.6.3}
\end{equation}

\underline{One can conclude} (it has obtained for the first time):
\begin{itemize}
\item
The Br probability in the first approximation at $l=0$, formed by the
Coulomb field both with taking into account of the nuclear forces of
any shape, and without such forces, does not depend on a value of the
angle $\beta$ between the directions of the $\alpha$-particle
propagation (or its tunneling) and the photon emission.

\item
The Coulomb field is degenerated by the quantum number $m$. This
property distinguishes the Coulomb field from the nuclear forces at
their account in the model. This difference is shown in the matrix
elements (\ref {eq.6.2}).
The nuclear forces participate in formation of the decay barrier and,
therefore, one can consider approximately them as forces working in
the spatial region of the barrier, where there is a tunneling.
One can assume, that one can divide the emissions from the barrier
region and from the external region on the basis of the quantum
number $m_{f}$.
It can be interesting to find a possible way of extraction of the Br
spectrum from the barrier region (or from the external region) from
experimental Br spectrum on the basis of this property.
\end{itemize}
% ***************************************************************************

% ***************************************************************************
\section{The second approximation at $l=1$
\label{sec.7}}

Taking into account (\ref{eq.6.1}), we find the Br probability in
the second approximation at $l=1$:
\begin{equation}
\begin{array}{lcl}
  p_{in}^{(l=1)}(w,\beta) & = &
    i\sqrt{3} \cos{\beta}
    \displaystyle\sum\limits_{\mu = -1, 1} h_{\mu}
      J_{in, m_{f}}(1,w), \\
  p_{c}^{(l=1)}(w,\beta) & = & \sqrt{6} \cos{\beta} J_{c}(1,w).
\end{array}
\label{eq.7.1}
\end{equation}

Neglecting by Br from the internal region, we obtain the following
expressions for the component of the Br probability of the second
approximation at $l=1$:
\begin{equation}
  W^{(l=1)}(w, \beta) =
    N_{0} k_{f} w \Bigl| p_{c}^{(l=1)}(w,\beta) \Bigr|^{2} =
    6 N_{0} k_{f} w \Bigl| J_{c}(1,w) \Bigr|^{2} \cos^{2}{\beta}
\label{eq.7.2}
\end{equation}
and for the total Br probability in the second approximation at
$l=1$:
\begin{equation}
\begin{array}{lr}
  W_{l=1}(w, \beta) =  W_{l=0}(w) \Bigl| 1 - N(w) \cos{\beta} \Bigr|^{2}, &
  N(w) = 3i \displaystyle\frac{J_{c}(1,w)}{J_{c}(0,w)}.
\end{array}
\label{eq.7.3}
\end{equation}

\underline{One can conclude} (it has found for the first time):
\begin{itemize}
\item
The dependence of the Br probability in the $\alpha$-decay on the
value of the angle $\beta$ between the directions of the
$\alpha$-particle propagation (or its tunneling) and the photon
emission has harmonic type (\ref{eq.7.3}).

\item
The account of the nuclear forces does not change the dependence of
the Br probability in the second approximation at $l=1$ on such angle
value.

\item
Exp.~(\ref{eq.7.3}) allows to find analytically maximums and
minimums in the Br spectra in dependence on the angle $\beta$.
\end{itemize}
% ***************************************************************************

% ***************************************************************************
\section{Convergence of calculations in asymptotic region 
\label{sec.8}}

There is an essential difficulty in the calculations of the Br
spectra for the given nucleus, concerned with obtaining of the radial
integrals (\ref{eq.5.2}) (or (\ref{eq.4.5})). This difficulty is
caused by that such integral is improper, and its sub-integral
function is oscillated and damped slowly with increasing of $r$.
The function damps weaker with increasing of $r$, the larger region
of integration should be taken into account in the numerical
integration.
For $^{210}\mbox{Po}$ the damping degree of the sub-integral function is
such as for reliable values of the first 2-3 digits for the Br
spectrum it needs to take into account (with the higher accuracy of
calculations) 1 million of oscillations of this function.

As an evident demonstration of this problem, let's consider an
one-dimensional integral:
\begin{equation}
    \int\limits^{+\infty}_{a} 
    \displaystyle\frac{\sin {x}}{x} dx.
\label{eq.8.1}
\end{equation}
An exact analytical value of this integral at $a=0$ is known from
theory of functions of complex variables, equal to $\pi /2$. The
numerical calculation of the integral (with use of simple method of
trapeziums, method of Gauss or other methods of the numerical
integration) allows to obtain quickly the same result also, but with
a given degree of accuracy (which determines a region of the
numerical integration).
It proves a convergence of the computer calculations of such integral
with a concrete choice of the parameter $a$.
But \emph{at weak increasing of the parameter $a$ the region of the
numerical integration for obtaining of the same calculating accuracy
for the integral (\ref{eq.8.1}) increases \underline{essentially},
and, therefore, a difficulty to calculate this integral numerically
increases essentially}.
However, the application of methods of theory of functions of complex
variables makes the calculation of such integral as simple again.
So, on the example of the simple integral (\ref{eq.8.1}) one can meet
with the numerical problem of the convergence of the calculations of
the improper integrals with the damping slowly, oscillating
sub-integral functions.

We fulfill an analysis of the convergence of the calculation of the
integral (\ref{eq.5.2}) on the basis of the analysis of damping of
its sub-integral function in the asymptotic region, which is defined
by wave functions in the initial $i$-, final $f$-states and the
spherical Bessel function of order $l$.

For enough large values of $r$ one can use an asymptotic
representation of the spherical Bessel function of order $l$:
\begin{equation}
  j_{l}^{(as)}(kr) = \displaystyle\frac{1}{kr}
    \sin{\biggl(kr - \displaystyle\frac{\pi l}{2} \biggl)}
\label{eq.8.2}
\end{equation}
or
\begin{equation}
\begin{array}{ccl}
  j_{2n}^{(as)}(kr) & = &
    (-1)^{n} j_{0}^{(as)}(kr) =
    (-1)^{n} \displaystyle\frac{\sin{kr}}{kr}, \\
  j_{2n+1}^{(as)}(kr) & = &
    (-1)^{n} j_{1}^{(as)}(kr) =
    (-1)^{n+1} \displaystyle\frac{\cos{kr}}{kr}.
\end{array}
\label{eq.8.3}
\end{equation}
where $n$ is a natural number.

The wave functions $\psi_{i}(r)$ and $\psi_{f}(r)$ of the initial
$i$- and the final $f$-states are linear combinations of the Coulomb
functions $F_{l}(\eta,\rho)$ and $G_{l}(\eta,\rho)$ (divided on
$\rho_{i,f}$, with quantum number $l=0$ or $l=1$ for the initial $i$-
or the final $f$-state, correspondingly). One can write the Coulomb
functions for $l$ in the asymptotic region so:
\begin{equation}
\begin{array}{cc}
  F_{l}(\eta, \rho) = \sin{\theta_{l}}, &
  G_{l}(\eta, \rho) = \cos{\theta_{l}}.
\end{array}
\label{eq.8.4}
\end{equation}
where
\begin{equation}
\begin{array}{lcllcl}
  \theta_{l} & = &
    \rho - \eta \log{2\rho} +
    \displaystyle\frac{1}{2}\pi l + \sigma_{l}(\eta), &
  \rho_{i,f} & = & k_{i,f} r, \\
  \sigma_{l} (\eta) & = & \arg \Gamma (i\eta+l+1), &
  \eta_{i,f} & = & \displaystyle\frac{m \nu}{k_{i,f}}.
\end{array}
\label{eq.8.5}
\end{equation}
where $\Gamma(x)$ is Gamma function with argument $x$, 
$\nu$ is Zomerfield parameter.

Now one can conclude:
\begin {itemize}
\item
The spherical Bessel function $j_{l}^{(as)}(kr)$ in the asymptotic
region damps (and oscillates) with increasing of $r$ equally for any
order $l$.

\item
The Coulomb functions $F_{0}(\eta_{i},\rho_{i})$ and
$G_{0}(\eta_{i},\rho_{i})$ of order $0$ for the initial $i$-state
and the Coulomb functions $F_{1}(\eta_{f},\rho_{f})$ and
$G_{1}(\eta_{f},\rho_{f})$ of order $l=1$ for the final $f$-state
damp in the asymptotic region with increasing of $r$ equally,
oscillate equally and are shifted at a phase between each other.

\item
The total sub-integral function of the integral (\ref{eq.5.2}) in the
asymptotic region damps with increasing of $r$ equally for any order
$l$.

\end {itemize}

Taking into account (\ref{eq.5.2}) and (\ref{eq.8.3}), we obtain:
\begin{equation}
\begin{array}{cc}
  J_{c}^{(as)}(2n,w)   = (-1)^{n} J_{c}^{(as)}(0,w), &
  J_{c}^{(as)}(2n+1,w) = (-1)^{n} J_{c}^{(as)}(1,w).
\end{array}
\label{eq.8.6}
\end{equation}
I.~e. one can reduce any integral inside the asymptotic region to
one of two integrals $J^{(as)}(0,w)$ or $J^{(as)}(1,w)$.

Let's find the matrix element $p_{c}^{(as)}(w,\beta)$ in the
asymptotic region:
\begin{equation}
\begin{array}{lcl}
  p_{c}^{(as)}(w,\beta) & = &
    i\sqrt{\displaystyle\frac{2}{3}} J_{c}^{(as)}(0,w)
    \Biggl( \displaystyle\sum\limits_{n=0}^{+\infty}
    (4n+1) P_{2n} (\cos{\beta}) \Biggr) + \\
    & + &
    \sqrt{\displaystyle\frac{2}{3}} J_{c}^{(as)}(1,w)
    \Biggl( \displaystyle\sum\limits_{n=0}^{+\infty}
    (4n+3) P_{2n+1} (\cos{\beta}) \Biggr).
\end{array}
\label{eq.8.7}
\end{equation}
Thus, we reduce the formula (\ref{eq.3.9}) for the Br spectra to
the linear combination of two radial integrals, which are convergent
(one can calculate them with a desirable accuracy limited by the
calculations accuracy of a concrete computer) and do not depend on
the angle, and factors - sums on $n$, into which the problem of
convergence is carried out (one can meet with it in (\ref{eq.3.9})).

It should seem, that one can cut off the region of the numerical
integration in one boundary $R$ for calculation of the integral
$J_{c}(l,w)$ from (\ref{eq.5.2}) for any $l$.
However, the calculation convergence of the integral is determined
not only by the damping of the sub-integral function at large $r$,
but by its behavior on the whole integration region also.
An analysis has shown, that the sub-integral function inside the
barrier region and inside the external region closer to the barrier
behaves so, that the calculation of the total integral becomes more
and more sensible to it with increasing of $l$ and the calculation
convergence becomes worse.
Therefore, \emph{for obtaining the reliable values of the integrals
$J_{c}(l,w)$ (for the same accuracy) it needs to increase the
external boundary $R$ of the integration region for larger $l$} (such
a conclusion has obtained by us is calculating of the angular Br
spectra for $^{210}\mbox{Po}$ also).
% ***************************************************************************

% ***************************************************************************
\section{Angular calculations for the Br spectra in the
$\alpha$-decay of $^{210}\mbox{Po}$
\label{sec.9}}

As a demonstration of the described above method, let's calculate
the angular Br spectra in the $\alpha$-decay of $^{210}\mbox{Po}$. For
a comparison of results obtained in such approach, with results
obtained by models
\cite{Papenbrock.1998.PRLTA,Tkalya.1999.PHRVA,Takigawa.1999.PHRVA},
we shall choose the potential parameters as
in~\cite{Maydanyuk.2003.PTP} (they coincide with the parameters of
the potential with the external Coulomb field
in~\cite{Tkalya.1999.PHRVA} and in~\cite{Papenbrock.1998.PRLTA}).

In spite of the fact that there are methods allowed to calculate
absolute values of the Br spectra, in this paper at first we shall
find the relative values of the Br spectrum for the given nucleus
and then we shall normalize obtained spectra at one selected point of
the Br experimental spectrum for the given angle value. This approach
as against the previous one allows with a larger accuracy to analyze a
behavior of the Br spectra in dependence on the angle (besides, it is
more easy in application).

In the beginning we calculate the total Br probability in the second
approximation at $l=1$ for the angle $90^{\circ}$ by (\ref{eq.6.3})
with taking into account (\ref{eq.6.3}) (because the component of the
Br probability (\ref{eq.7.2}) in the second approximation at $l=1$
equals to zero at such angle). Then we normalize the obtained spectrum
by the third point of the experimental data \cite{D'Arrigo.1994.PHLTA}
(we have such values $w=0.179$ keV and
$W = 10.1 \cdot 10^{-10}$ 1 / keV / decay), which were obtained for
the angle $\beta = 90^{\circ}$ also. Knowing the normalized factor
and using formulas (\ref{eq.7.3}), we find the Br probability in the
second approximation for the other values of the angle $\beta$.

The angular values of the Br probability in the second approximation
at $l=1$ are shown in the Table 1. Here, one can see a variation of
the Br probability in dependence on the angle $\beta$, however this
change is extremely small.
The Br probability in the first approximation at $l=0$ coincides with
the Br probability in the second approximation at $l=1$ for the angle
$90^{\circ}$.
One can see that a contribution of the Br probability in the first
approximation into the total spectrum is the largest for any angle
value, i.~e. it is extremely larger then the contribution of the
component of the Br probability in the second approximation at $l=1$
into the total spectrum.
\emph{This conclusion has a physical sense (obtained for the first
time):
Br in the $\alpha$-decay for $^{210}\mbox{Po}$ depends extremely weakly
on the value of the angle between directions of the $\alpha$-particle
propagation (or tunneling) and the photon emission (in the given
approach)}.
The account of not zero component of the Br probability in the second
approximation (for the angle values which are distinct from
$90^{\circ}$) increases the total Br probability.
Absolute and relative variations of the Br probability relatively its
maximal and minimal values
\begin{equation}
\begin{array}{lcl}
  \Delta W_{1}(w) & = &
    W_{l=1}(w, \beta=0^{\circ}) - W_{l=1}(w, \beta=90^{\circ}); \\
  \Delta W_{2}(w) & = &
    \displaystyle\frac
    {|W_{l=1}(w, \beta=0^{\circ}) - W_{l=1}(w, \beta=90^{\circ})|}
    {W_{l=1}(w, \beta=90^{\circ})} \cdot 100
\end{array}
\label{eq.9.1}
\end{equation}
are included into the table also.
\begin{table}
% \hspace{-20mm}
\begin{center}
\begin{tabular}{|c|c|c|c|c|c|c|c|c|c|} \hline
 $w$, &
 \multicolumn{7}{|c|}{$W_{l=1}(w, \beta)$, 1 / keV / decay} &
 $\Delta W_{1}$, &
 $\Delta W_{2}$ \\ \cline{2-8}
  keV &
  $\beta=0^{\circ}$ &
  $\beta=15^{\circ}$ &
  $\beta=30^{\circ}$ &
  $\beta=45^{\circ}$ &
  $\beta=60^{\circ}$ &
  $\beta=75^{\circ}$ &
  $\beta=90^{\circ}$ &
  1/keV/dec. &
    \\ \hline
   50 &  1.641E-08 &  1.635E-08 &  1.617E-08 &  1.589E-08 &  1.553E-08 &  1.511E-08 &  1.467E-08 &  1.735E-09 &  11.8 \\
  100 &  4.974E-09 &  4.953E-09 &  4.892E-09 &  4.796E-09 &  4.673E-09 &  4.531E-09 &  4.381E-09 &  5.928E-10 &  13.5 \\
  150 &  1.897E-09 &  1.890E-09 &  1.869E-09 &  1.836E-09 &  1.793E-09 &  1.744E-09 &  1.692E-09 &  2.047E-10 &  12.1 \\
  200 &  8.021E-10 &  7.993E-10 &  7.912E-10 &  7.783E-10 &  7.618E-10 &  7.427E-10 &  7.226E-10 &  7.949E-11 &  11.0 \\
  250 &  3.548E-10 &  3.534E-10 &  3.493E-10 &  3.429E-10 &  3.346E-10 &  3.250E-10 &  3.149E-10 &  3.993E-11 &  12.7 \\
  300 &  1.611E-10 &  1.605E-10 &  1.585E-10 &  1.554E-10 &  1.515E-10 &  1.469E-10 &  1.421E-10 &  1.901E-11 &  13.4 \\
  350 &  7.628E-11 &  7.590E-11 &  7.480E-11 &  7.306E-11 &  7.082E-11 &  6.826E-11 &  6.557E-11 &  1.071E-11 &  16.3 \\
  400 &  3.251E-11 &  3.236E-11 &  3.194E-11 &  3.127E-11 &  3.042E-11 &  2.943E-11 &  2.840E-11 &  4.110E-12 &  14.5 \\
  450 &  1.278E-11 &  1.275E-11 &  1.266E-11 &  1.252E-11 &  1.234E-11 &  1.213E-11 &  1.191E-11 &  8.716E-13 &   7.3 \\
  500 &  6.094E-12 &  6.077E-12 &  6.030E-12 &  5.956E-12 &  5.860E-12 &  5.749E-12 &  5.632E-12 &  4.615E-13 &   8.2 \\
  550 &  3.198E-12 &  3.179E-12 &  3.124E-12 &  3.038E-12 &  2.928E-12 &  2.802E-12 &  2.671E-12 &  5.271E-13 &  19.7 \\
  600 &  1.624E-12 &  1.612E-12 &  1.578E-12 &  1.524E-12 &  1.455E-12 &  1.378E-12 &  1.297E-12 &  3.266E-13 &  25.2 \\
  650 &  5.731E-13 &  5.712E-13 &  5.656E-13 &  5.569E-13 &  5.456E-13 &  5.326E-13 &  5.189E-13 &  5.422E-14 &  10.5 \\
  700 &  2.198E-13 &  2.186E-13 &  2.150E-13 &  2.094E-13 &  2.023E-13 &  1.942E-13 &  1.858E-13 &  3.398E-14 &  18.3 \\
  750 &  9.515E-14 &  9.445E-14 &  9.240E-14 &  8.920E-14 &  8.512E-14 &  8.050E-14 &  7.571E-14 &  1.944E-14 &  25.7 \\
  800 &  2.409E-14 &  2.411E-14 &  2.418E-14 &  2.430E-14 &  2.446E-14 &  2.467E-14 &  2.490E-14 & -8.094E-16 &   3.3 \\ \hline
\end{tabular}
\end{center}
\caption{Angular values of Br probability in $\alpha$-decay of
$^{210}\mbox{Po}$ in approximation $l=1$
\label{table.1}}
\end{table}

\begin{figure}[h]
\centering\includegraphics[width=11cm]{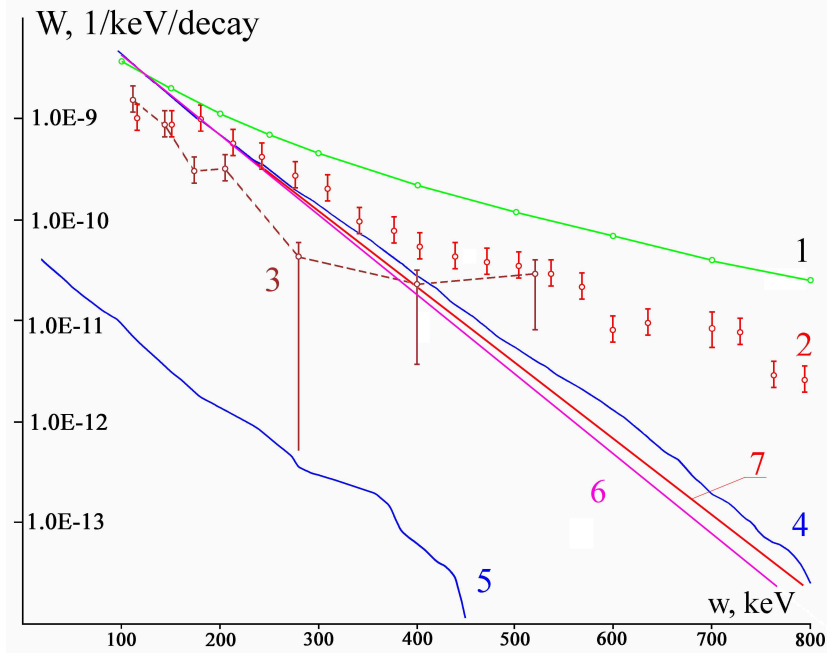}
% \caption{Спектры ТИ при распаде $^{210}\mbox{Po}$:
% 1 --- кривая, полученная в подходе модели мгновенного ускорения
% (на основе классической элекстродинамики) и извлеченная из
% \cite{Tkalya.1999.PHRVA};
% 2 --- экспериментальные данные из \cite{D'Arrigo.1994.PHLTA,Eremin.2000.PRLTA};
% 3 --- экспериментальные данные из \cite{Kasagi.1997.PRLTA};
% 4 --- кривая вероятности ТИ $W_{l=1}$ в первом приближении $l=1$ по
% нашему методу, практически совпадает с кривыми по квантово-механическим
% моделям \cite{Papenbrock.1998.PRLTA} и \cite{Tkalya.1999.PHRVA};
% 5 --- кривая составляющей $W^{(l=1)}$ вероятности ТИ первого приближения
% $l=1$ для угла $45^{\circ}$ по нашему методу.
%
\caption{The Br spectra in $\alpha$-decay of $^{210}\mbox{Po}$:
1 is a curve, extracted from \cite{Tkalya.1999.PHRVA} by the instant
accelerated model;
2 is the experimental data \cite{D'Arrigo.1994.PHLTA,Eremin.2000.PRLTA};
3 is the experimental data \cite{Kasagi.1997.PRLTA};
4 is a curve of the Br probability $W_{l=1}$ in the first approximation
$l=1$ by our approach at $\beta=90^{\circ}$;
5 is a curve of the Br probability component $W^{(l=1)}$ in the
first approximation $l=1$ at the angle $45^{\circ}$ by our approach;
6 is a curve calculated by us with radial integral (6) and formula
$d\psi_{i}(r) / dr = - \psi_{i}(r)/w \: dV(r) / dr$
(it coincides with radial integral (7) in \cite{Tkalya.1999.PHRVA}
with factor $-1/w$)
and further normalization at the third point of data
\cite{D'Arrigo.1994.PHLTA,Eremin.2000.PRLTA};
7 is a curve calculated on the basis of the potential
(\ref{eq.11.1})--(\ref{eq.11.3})
\label{fig.1}}
\end{figure}

The results of the calculations of the Br probability in the
$\alpha$-decay of $^{210}\mbox{Po}$ in the second approximation at $l=1$
by our approach are shown in Fig.~\ref{fig.1}.
From the figure one can see, that for the angle $90^{\circ}$ our
method gives the Br spectrum, which is very close to the Br spectra,
obtained by the models \cite{Papenbrock.1998.PRLTA} and
\cite{Tkalya.1999.PHRVA}. However, as against the models
\cite{Papenbrock.1998.PRLTA,Tkalya.1999.PHRVA} our method shows the
angular variation of the Br spectra.
% ***************************************************************************

% ***************************************************************************
\section{An analyses of the maximums and minimums in the angular Br
spectra
\label{sec.10}}

Let's find the values of the angle $\beta$ between the directions of
the $\alpha$-particle propagation (or tunneling) and the photon
emission, at which the Br probability has the maximal and minimal
values. Using a derivative
\begin{equation}
  \displaystyle\frac{d W_{l=1}(w, \beta)}{d\beta} =
    W_{l=0}(w)
    \Bigl(N^{*}(w) + N(w) - 2|N(w)|^{2}\cos{\beta} \Bigr) \sin{\beta},
\label{eq.10.1}
\end{equation}
we find conditions of extremal values of the function
$W_{l=1}(w,\beta)$:
\begin{equation}
\begin{array}{l}
  \sin(\beta) = 0; \\
  \cos(\beta) =
    \displaystyle\frac{N^{*}(w) + N(w)} {2|N(w)|^{2}} =
    \displaystyle\frac{Re(N(w))} {Re(N(w))^{2} + Im(N(w))^{2}}.
\end{array}
\label{eq.10.2}
\end{equation}

Calculations for $^{210}\mbox{Po}$ for the given potential have shown,
that the second condition in (\ref{eq.10.2}) in a range $w$ = 50--800
keV is not carried out. One can explain this by that the integral
$J_{c}(1,w)$ is smaller than the integral $J_{c}(0,w)$ in
$10^{2}$--$10^{4}$ times (that is in agreement with a condition of
the convergence of the Br spectra at increasing of $l$).
From the first condition in (\ref{eq.10.2}) we obtain such extremal
values for the angle $\beta$:
\begin{equation}
  \beta = 0, \pi.
\label{eq.10.3}
\end{equation}
The Br probability with such angle values has the maximal and
minimal values, and between them it varies monotonously for any
energy of the photon emitted in the range $w$ = 50--800 keV. One can
see this also from the Table 1.
% ***************************************************************************

% ***************************************************************************
\section{Inclusion of Woods-Saxon potential into the model 
\label{sec.11}}

Now let's analyze, how the Br spectrum in the $\alpha$-decay of the
studied nucleus $^{210}\mbox{Po}$ is changed, if in the approach for
calculations of the Br spectra from the potential of interaction
between the $\alpha$-particle and the daughter nucleus, pointed out in
Sec.~\ref{sec.9} and having the simplified barrier, to pass to a
potential with the barrier, constructed on the basis of the account
of realistic nuclear forces of interaction between the
$\alpha$-particle and the daughter nucleus which is used in realistic
nuclear models.

With such a purpose we shall take a potential, proposed
in~\cite{Denisov.2005.PHRVA} for description of the $\alpha$-decay
and synthesis of nuclei. Among extensive set of literature giving us
different types of the $\alpha$-nucleus potentials, we have given
preference to such a paper because of here we see universal and clear
approach for calculation of parameters of the potential after choice
of needed such nucleus. In result, we suppose to obtain universal
recipe for calculation of the Br spectra after choosing the nucleus.

So, according to~\cite{Denisov.2005.PHRVA} (see~(6)--(10)), we use
such potential of interaction:
\begin{equation}
  V (r, \theta, l, Q) =
    V_{C} (r, \theta) + V_{N} (r, \theta, Q) + V_{l} (r),
\label{eq.11.1}
\end{equation}
where
\begin{equation}
\begin{array}{lcl}
  V_{C} (r, \theta) & = &
    \displaystyle\frac{2 Z e^{2}} {r}
    \biggl(1 +
    \displaystyle\frac{3 R^{2}} {5 r^{2}} \beta Y_{20}(\theta) \biggr), \\

  V_{N} (r, \theta, Q) & = &
    \displaystyle\frac{v(A,Z,Q)}
    {1 + \exp{\displaystyle\frac{r-r_{0}} {d}}}, \\

  V_{l} (r) & = & \displaystyle\frac{l(l+1)} {2mr^{2}}.
\end{array}
\label{eq.11.2}
\end{equation}
At current stage (with a purpose to simplify the numerical
calculations of the Br spectra), for determination of the component
$V_{C}(r,\theta)$ we use the formula (7) in~\cite{Denisov.2005.PHRVA}
on whole region of $r$
(without use of (8) in~\cite{Denisov.2005.PHRVA}).

According to (14)--(20) in~\cite{Denisov.2005.PHRVA}, we calculate
the parameters as follows:
\begin{equation}
\begin{array}{rcl}
  v(A,Z,Q) & = &
    -(30.275 - 0.45838 Z/A^{1/3} + 58.270 I - 0.24244 Q), \\
  r_{m} & = & 1.5268 + R, \\
  R & = & R_{p}(1 + 3.0909/R_{p}^{2}) + 0.1243 t, \\
  R_{p} & = & 1.24 A^{1/3} (1 + 1.646/A - 0.191 I), \\
  t & = & I - 0.4 A/(A+200), \\
  d & = & 0.49290,
\end{array}
\label{eq.11.3}
\end{equation}

According to \cite{Muntyan.2003}, we see, that the parameter $\beta$
for $^{210}\mbox{Po}$ is very small, that points out to a high degree
of sphericity of this nucleus. Therefore, for calculation of the Br
spectra we note the following:
\begin {itemize}

\item
In definition of $r_{m}(\theta)$ and $R(\theta)$ we do not use (21)
and (22) of~\cite{Denisov.2005.PHRVA} (using (15)
from~\cite{Denisov.2005.PHRVA}).

\item
The formalism for calculation of the Br spectra, presented in 
Sec.~\ref{sec.3}--\ref{sec.10}), is constructed on the basis of
division of the total wave function into its radial and angular
components, i.~e. in the assumption of spherical symmetry of the
decaying nucleus.
\emph{Therefore, for the nucleus $^{210}\mbox{Po}$ our approach for
calculation of the Br spectra is applicable with account of realistic
nuclear forces also}.
\end {itemize}  

Further, we calculate the radial wave function of the decaying system
for the potential (\ref{eq.11.1})--(\ref{eq.11.3}).
This gives us the general solution for the wave function in
dependence on the selected energy level for $\alpha$-decay.
To achieve, that the found solutions describe the states of the
decaying system before and after the spontaneous photon emission, we
should take into account boundary conditions in initial and final
states.
Here, we use such conditions:
\begin {equation}
\begin {array} {ll}
  \mbox {the initial $i$-state:} & \chi_{i}(r \to +\infty) \to G(r)+iF(r), \\
      \mbox {the final $f$-state:} & \chi _ {f} (r=0) = 0,
\end {array}
\label{eq.11.4}
\end {equation}
where
$\varphi_{i,f}(r) = \displaystyle\frac{\chi_{i,f}(r)}{r}$,
$F$ and $G$ are Coulomb functions.

One note, that in against of a scattering of the $\alpha$-particle
on the nucleus, where as the boundary condition for the initial
$i$-state \emph{a finiteness of the radial wave function
$\varphi_{i}(r)$ should be used at point $r=0$ ($\chi_{i}(r=0)=0$)}, 
for decay we choose a natural requirement, that the radial wave
function tend to a spherical divergent wave in asymptotic region
($\chi_{i}(r\to\infty)$ tend to a plane divergent wave).
\emph{This condition gives us inevitably the divergence of the total
radial wave function in the initial $i$-state at point $r=0$
(which real and imaginary parts consist from regular and singular
solutions)!}
One can make sure in this by requiring a fulfillment of the continuity
condition for the radial wave function on the whole region on its
definition; or by requiring the constancy of the radial flux density,
which is distinct from zero and is directed outside in the asymptotic
region (and, therefore, it should be not zero near point $r=0$, that
is impossible to execute with null wave function at any chosen point
of $r$).
This peculiarity complicates essentially the calculations of the Br
spectra for the $\alpha$-decay in comparison with the problems of the
calculation of the Br spectra for the scattering of charged particles
on nuclei (where the essential progress has been achieved and a lot of
papers are published).

\emph{It turns out, that real and imaginary parts of a sub-integral
function of (\ref {eq.4.5}) for the calculation of the matrix elements,
constructed on the basis of the found solutions for the wave function
for the initial and final states, tend to zero at point $r\to 0$!}
This interesting peculiarity provides the convergence of the matrix
elements near point $r=0$ and, therefore, in the whole region of $r$
(in the asymptotic region the convergence of the wave function is
determined by the convergence of the Coulomb functions, considered
above).
Thus, we resolve the divergence problem in the calculations of the
Br spectra in the $\alpha$-decay of the nucleus $^{210}\mbox{Po}$.

The Br spectrum for the nucleus $^{210}\mbox{Po}$ with the
$\alpha$-nucleus potential (\ref{eq.11.1})--(\ref{eq.11.3}) by our
approach and calculations is shown by the curve with number 7 in
Fig.~\ref{fig.1}.
From here one can see, that the new curve 7 of the Br probability is
located very close to the curve 4 for the Br probability with the
potential from Sec.~\ref{sec.9} with the simplified barrier (and also
close to the curve 6 by the approaches
\cite{Papenbrock.1998.PRLTA,Tkalya.1999.PHRVA}).

\vspace {5mm}
\noindent
\underline {Conclusions:}
\begin {itemize}
\item
The account of nuclear forces, essentially changed a shape of the
barrier in its internal part (and essentially changed Br from this
internal region), changes very slowly the spectrum of the total Br
in the $\alpha$-decay of the nucleus $^{210}\mbox{Po}$ (in comparison
with the earlier obtained Br spectrum on the basis of the potential
from Sec.~\ref{sec.9} with the simplified barrier).

\item
This point confirms the result (obtained early on the basis of the
$\alpha$-nucleus potential pointed out in Sec.~\ref{sec.9} with the
barrier of the simplified shape) that the Br emission from the
internal region till point $r$ for barrier maximum gives the very
small contribution into the total Br spectrum.
This conclusion coincides logically with a property (found on the
basis of microscopic models of the nuclei with their $\alpha$-decay)
of leaving of the $\alpha$-particle from the nuclear surface during
the first decay stage.

\end {itemize}
% ***************************************************************************

% ***************************************************************************
\section{Conclusions and perspectives 
\label{conclusions}}

We present the new method of calculation of the Br spectra in the
$\alpha$-decay, where the angle between the directions of the
$\alpha$-particle motion (with tunneling) and the photon emission
(used in the experiments \cite{Eremin.2000.PRLTA,Kasagi.2000.PRLTA})
is taken into account. Using it, the angular spectra for the nucleus
$^{210}\mbox{Po}$ are obtained for the first time. Now let us formulate
the main conclusions and perspectives:

\begin{itemize}
\item
The method gives such a dependence of the bremsstrahlung spectrum in
the $\alpha$-decay of $^{210}\mbox{Po}$ on the angle (this has been obtained
for the first time):
\begin{itemize}
\item
the first approximation at $l=0$ gives independence of the spectrum
on the angle;

\item
the second approximation at $l=1$ gives a slow monotonous variation
of the slope of the spectrum curve with changing the angle and without
a visible change of the shape of the spectrum curve (i.~e. without the
appearance of humps and holes in the spectrum);

\item
for arbitrary energy of the photon emitted in the range of
$w$ = 50--750 keV the bremsstrahlung probability is maximal at the
angle $0^{\circ}$ and is minimal at the angle $180^{\circ}$, between
these angular values the bremsstrahlung probability varies
monotonously.
\end{itemize}

\item
Results for $^{210}\mbox{Po}$ have obtained on the basis of these
approximations:
\begin{itemize}

\item
the bremsstrahlung process does not depend on the direction of the
leaving the $\alpha$-particle relatively to the shape of the daughter
nucleus before the photon emission (this supposition has been fulfilled
for $^{210}\mbox{Po}$, because, in accordance with \cite{Muntyan.2003}
(see Fig.~5 on p.~33), coefficients $\beta_{\lambda}^{0}$ of the
shape deformation for this nucleus at $\lambda=2,4,6,8$ are extremely
close to zero in comparison with other nuclei with other numbers of
protons and neutrons, i.~e. $^{210}\mbox{Po}$ is one of the most
spherical nuclei);

\item
the photon emission does not change the direction of the
$\alpha$-particle propagation (this supposition is suitable for the
low-energy photons, and, therefore, it is applicable for analysis of
all existing experimental data of the bremsstrahlung spectra, where
one can select a region with smaller photons energies for increasing
accuracy);

\item
the bremsstrahlung spectra have been calculated by means of the 
$\alpha$-nucleus potential with the simplified barrier pointed out in 
Sec.~\ref{sec.9} (they coincide with the $\alpha$-nucleus potential 
in~\cite{Maydanyuk.2003.PTP}, and also with the potential with the
external Coulomb field in~\cite{Tkalya.1999.PHRVA} and
in~\cite{Papenbrock.1998.PRLTA}) and with use of the $\alpha$-nucleus
potential with the barrier, pointed out in Sec.~\ref{sec.11}
(see~\cite{Denisov.2005.PHRVA}) and constructed on the basis of
realistic nuclear forces.
\end{itemize}

\item
Taking into account nuclear forces in the method gives the following:
\begin{itemize}
\item
it does not change dependences on the angle of bremsstrahlung
probability in the first and second approximations;

\item
it essentially changes the shape of the barrier in its internal region
(sufficiently changes the Br from such internal region) and changes
very little the spectrum of the total Br in the $\alpha$-decay
for the nucleus $^{210}\mbox{Po}$ at selected angle value.

\end{itemize}
\end{itemize}

From here a question naturally arises: 
\emph{which improvement should be made in the method that results
in enough visible changes of the Br spectrum curve, to achieve a
better description of the experimental data?} Note the following:
\begin {itemize}

\item
According to our calculations, in consideration of the possibility of
the $\alpha$-particle leaving at the energy of the \emph{exited state} of
the decaying system, the angle of the slope of the Br spectrum curve
increases (monotonously).
Apparently, it allows to displace the calculated Br curve (for example,
by our method) essentially closer to the experimental data
\cite{D'Arrigo.1994.PHLTA}.

\item
For obtaining \underline{reliable values} of the Br spectra for the 
$\alpha$-decay one needs to achieve in the calculations the convergence
of integrals for such spectra.
This leads to the necessity to consider wave functions inside a large
space region with the external boundary far enough from the nucleus.
From here a new question naturally arises as to taking into 
account electrons shells of the atom with such nuclear
$\alpha$-decay in the calculation of the Br spectra in the
$\alpha$-decay and one can formulate the following \emph{hypothesis
about the visible influence of the electrons shells of the atom on the
total Br spectrum in the $\alpha$-decay}.
One can note that an essential progress has been made by
M.~Amusia in the study of the Br in atomic physics
\cite{Amusia.1988.PRPLC,Amusia.1990}.
Moreover, according to \cite{Amusia.2005} (see~p.20--21), there is an
inevitable influence of the $\alpha$-decay process at its starting
time stage on the electrons shells of the atom whose nucleus decays.
So, the $\alpha$-particle during its leaving (with tunneling) deforms
and polarizes these electrons shells.
In one's turn, the changed electrons shells can correct our
understanding of the real $\alpha$-nucleus potential in the model,
which should be used for the calculations of the Br spectra
(one should note that these effects are still not studied in details).
Note that these effects (partially) take place in the same space region,
where it is necessary to use the wave functions for the calculation
of the matrix elements to achieve converging values of the Br
spectra.
Therefore, it is desirable to study these effects to obtain for more
accurately the total Br spectra of the $\alpha$-decay.

\item
The inclusion of the $\alpha$-nucleus potential from
\cite{Denisov.2005.PHRVA} in our method, deforming the decay
barrier, does not essentially displace a point, where the $\alpha$-particle
starts to tunnel through the barrier.
It turns out that the displacement of this point is much smaller in
comparison with the tunneling region and even with a \emph{``mixed
region''}, introduced in \cite{Takigawa.1999.PHRVA}.
Therefore, after taking into account the realistic nuclear forces
in the method, the interest to analyze the Br from these regions
(with detailed study of tunneling) remains.

\item
We assume that further development of the time formalism for the
description of the Br in the $\alpha$-decay at its first stage
will give new abilities in the accurate description of the Br. One
cannot exclude the assumption about appearance of \emph {``holes''} in the
Br spectra (see \cite{Kasagi.1997.JPHGB,Takigawa.1999.PHRVA}), that
can allow to better describe the experimental data
\cite{Kasagi.1997.JPHGB,Kasagi.1997.PRLTA}. However, in such a case
it is not clear how to connect this with the available experimental data
\cite{D'Arrigo.1994.PHLTA} without of ``holes'' in the logical basis
of our method.
\end{itemize}

Now let's formulate the conclusions, which have a physical sense and
on the basis of the calculations for $^{210}\mbox{Po}$ by our model:
\begin{itemize}
\item
The bremsstrahlung in the $\alpha$-decay of the spherical nuclei
depends on the angle extremely weakly. In taking into account of the
nuclear forces, such dependence is not changed visibly.

\item
It is not enough to take into account only one angle for the
explanation of the difference between the experimental spectra
\cite{D'Arrigo.1994.PHLTA} and
\cite{Kasagi.1997.JPHGB,Kasagi.1997.PRLTA} for $^{210}\mbox{Po}$
(which equals to $90^{\circ}$ and $25^{\circ}$, correspondingly)
on the basis of our model
and for the explanation of the difference between these experimental 
spectra and the calculated curves averaged by angle values in the
approaches
\cite{Papenbrock.1998.PRLTA,Takigawa.1999.PHRVA,Tkalya.1999.PHRVA}
(that can be supposed in~\cite{Eremin.2000.PRLTA,Kasagi.2000.PRLTA}).

\item
The small visible change of the Br spectra after taking into account
of the realistic nuclear forces in our method confirms the result
(obtained on the basis of the $\alpha$-nucleus potential from
Sec.~\ref{sec.9} with the barrier of the simplified shape, also see
\cite{Maydanyuk.2003.PTP}) that the Br from the internal region till
point $r$ for barrier maximum gives the very small contribution into
the total Br spectrum. This conclusion become natural if to take into
account such a property (found on the basis of microscopic models of
the $\alpha$-decay) as the $\alpha$-decay starts from leaving of the
$\alpha$-particle from the nuclear surface.

\end{itemize}

In closing, supposing that the Br spectra in the $\alpha$-decay must
to change essentially at change of the angle value, we note, how this
point can be explained, analyzing this question in theoretical and
experimental aspects:
\begin{itemize}

\item
One can explain such angular change of the bremsstrahlung spectrum so:
\begin{itemize}

\item
In the $\alpha$-decay of the (initially) spherical nuclei --- 
by strong angular deformation of the decay barrier and continuous 
redistribution of the electromagnetic charge (or \emph{``nuclear
polarization''} likely the polarization of the electrons shells
during tunneling of the $\alpha$-particle, according to
\cite{Amusia.2005} (see p.~20--21)).
One can suppose, that here non-central forces between the
$\alpha$-particle and nucleons of the daughter nucleus play essential
role, which exist in the barrier region mainly.
Note, that a serious progress was achieved early in the microscopic
approach in study of the bremsstrahlung in scattering of the
nucleons and the $\alpha$-particles on the light nuclei (see
\cite{Baye.1985.NUPHA,Descouvemont.1986.PHLTA,%
Liu.1990.PHRVA.C41,Liu.1990.PHRVA.C42,Baye.1992.NUPHA}),
in study of the bremsstrahlung in collisions between heavy ions
and the nuclei (see \cite{Biro.1987.NUPHA}),
and in the non-microscopic approaches in study of bremsstrahlung
induced by protons during their collisions on heavier nuclei
(see \cite{Pluyko.1987.FECAA}).

\item
In the $\alpha$-decay of the deformed nuclei --- by essential
appearance of the angular anisotropy of the $\alpha$-nucleus potential.
Then one can extract an information about the shape of the nucleus
from the angular bremsstrahlung spectra.

\end{itemize}

\item
Experimental confirmation of the change the Br spectrum in the
$\alpha$-decay of the spherical nuclei at change of the angle value
gives the following:
\begin{itemize}

\item
It will prove an existence of essential microscopic forces between
the $\alpha$-particle and the nucleons of the daughter nucleus,
reinforcing the angular deformation of the barrier.

\item
It will prove a visible influence of the bremsstrahlung on dynamics
of the $\alpha$-decay. In accordance with our model, it will be an
experimental confirmation of the effect of variation of the barrier
penetrability in result of the emission during tunneling of a charged
particle, proposed in \cite{Flambaum.1999.PRLTA}.

\end{itemize}
\end{itemize}

The angular analysis of the bremsstrahlung spectra gives the new
additional information about the $\alpha$-decay. Therefore, further
angular experimental measurements of the bremsstrahlung spectra will
have the interest.
% ***************************************************************************

% ***************************************************************************
\section*{Acknowledgements
\label{acknowledgements}}

Authors express their deep gratitude to Dr.~I.~E.~Kashuba for his
valuable assistance in development of numerical algorithms for
calculation of Coulomb functions with higher accuracy, that has
allowed to obtain a convergence in calculations of the
bremsstrahlung spectra for $^{210}\mbox{Po}$.
% ***************************************************************************

% ***************************************************************************
% \bibliographystyle{h-physrev4}
\bibliography{Be}

\begin{thebibliography}{}

\bibitem{D'Arrigo.1994.PHLTA}
  A.~D'Arrigo, N.~V.~Eremin, G.~Fazio, G.~Giardina, M.~G.~Glotova,
  T.~V.~Klochko, M.~Sacchi and A.~Taccone,
\newblock
  {\it Investigation of bremsstrahlung emission in $\alpha$-decay
  of heavy nuclei},
\newblock
  Physics Letters \textbf{B332} (1--2), 25--30 (July, 1994).
%%CITATION = PHLTA,B332,25;%%

\bibitem{Kasagi.1997.JPHGB}
  J.~Kasagi, H.~Yamazaki, N.~Kasajima, T.~Ohtsuki and H.~Yuki,
\newblock
  {\it Bremsstrahlung emission in $\alpha$-decay and tunneling motion
  of $\alpha$-particle},
\newblock
  Journal of Physics G: Nuclear and Particle Physics {\bf 23},
  1451--1457 (1997).
%%CITATION = JPHGB,23,1451;%%

\bibitem{Kasagi.1997.PRLTA}
  J.~Kasagi, H.~Yamazaki, N. Kasajima, T.~Ohtsuki and H.~Yuki,
\newblock
  {\em Bremsstrahlung in $\alpha$-decay of $^{210}\mbox{Po}$:
  do $\alpha$-particles emit photons in tunneling?}
\newblock
  Physical Review Letters {\bf 79} (3), 371--374 (July, 1997).
%%CITATION = PRLTA,79,371;%%

\bibitem{Eremin.2000.PRLTA}
  N.~V.~Eremin, G.~Fazio and G.~Giardina,
\newblock
  {\it Comment on ``Bremsstrahlung in $\alpha$-decay of
  $^{210}\mbox{Po}$: do $\alpha$-particles emit photons in tunneling?''}
\newblock
  Physical Review Letters {\bf 85} (14), 3061 (October, 2000).
%%CITATION = PRLTA,85,3061;%%

\bibitem{Kasagi.2000.PRLTA}
  J.~Kasagi, H.~Yamazaki, N.~Kasajima, T.~Ohtsuki and H.~Yuki,
\newblock
  {\it Replay on Comment on ``Bremsstrahlung in $\alpha$-decay of
  $^{210}\mbox{Po}$: do $\alpha$-particles emit photons in tunneling?''}
\newblock
  Physical Review Letters {\bf 85} (14), 3062 (October, 2000).
%%CITATION = PRLTA,85,3062;%%


\bibitem{Batkin.1986.SJNCA}
%   И.~С.~Баткин, И.~В.~Копытин, Т.~А.~Чуракова
% \newblock
%   {\it Внутреннее тормозное излучение, сопровождающее $\alpha$-распад},
% \newblock
%   Ядерная физика, {\bf т.~44}, вып.~6 (12), 1454--1458 (1986).
%
  I.~S.~Batkin, I.~V.~Kopytin and T.~A.~Churakova
\newblock
  {\it Internal bremsstrahlung accompanying $\alpha$ decay},
\newblock
  Yad. Phys. (Sov. Journal of nuclear physics) {\bf Vol.~44},
  Iss.~6 (12), 1454--1458 (1986).

\bibitem{Papenbrock.1998.PRLTA}
  T.~Papenbrock and G.~F.~Bertsch,
\newblock
  {\it Bremsstrahlung in $\alpha$-decay},
\newblock
  Physical Review Letters {\bf 80} (19), 4141--4144 (May, 1998);
\newblock
  [arXiv:nucl-th/9801044].
%%CITATION = NUCL-TH 9801044;%%

\bibitem{Takigawa.1999.PHRVA}
  N.~Takigawa, Y.~Nozawa, K.~Hagino, A.~Ono and D.~M.~Brink,
\newblock
  {\it Bremsstrahlung in $\alpha$ decay''},
\newblock
  Physical Review {\bf C59} (2), 593--597 (February, 1999);
\newblock
  [nucl-th/9809001].
%%CITATION = NUCL-TH 9809001;%%

\bibitem{Tkalya.1999.PHRVA}
  E.~V.~Tkalya,
\newblock
  {\it Bremsstrahlung in $\alpha$-decay and ``interference of
  space regions''},
\newblock
  Physical Review {\bf C60} (5), 054612--054615 (1999).
%%CITATION = PHRVA,C60,054612;%%

\bibitem{Kurgalin.2004}
  S.~D.~Kurgalin, Yu.~M.~Chuvilsky and T.~A.~Churakova,
\newblock
  {\it Modelirovanie harakteristik tormoznogo $\gamma$-izlucheniya v
  $\alpha$-raspadah $^{226}\mbox{Ra}$ i $^{214}\mbox{Po}$},
\newblock
  Vestnik VGU, Seria fizika, matematika, {\bf 1}, 21--26 (2004).

% \bibitem{Kurgalin.2004}
%   С.~Д.~Кургалин, Ю.~М.~Чувильский, Т.~А.~Чуракова,
% \newblock
%   {\it Моделирование характеристик тормозного $\gamma$-излучения в
%   $\alpha$-распадах $^{226}\mbox{Ra}$ и $^{214}\mbox{Po}$},
% \newblock
%   Вестник ВГУ, Серия физика, математика, {\bf 1}, 21--26 (2004).

\bibitem{Dyakonov.1996.PRLTA}
  M.~I.~Dyakonov and I.~V.~Gornyi,
\newblock
  {\it Electromagnetic radiation by a tunneling charge},
\newblock
  Physical Review Letters {\bf 76} (19), 3542--3545 (May, 1996).
%%CITATION = PRLTA,76,3542;%%

\bibitem{Dyakonov.1999.PHRVA}
  M.~I.~Dyakonov,
\newblock
  {\it Bremsstrahlung spectrum in $\alpha$ decay},
\newblock
  Physical Review {\bf C60}, 037602-4p. (July, 1999);
\newblock
  [arXiv:nucl-th/9903016].
%%CITATION = NUCL-TH 9903016;%%

\bibitem{Bertulani.1999.PHRVA}
  C.~A.~Bertulani, D.~T.~de~Paula and V.~G.~Zelevinsky,
\newblock
  {\it Bremsstrahlung radiation by a tunneling particle:
  A time-dependent description},
\newblock
  Physical Review {\bf C60} (3), 031602--4p. (August, 1999);
\newblock
  [arXiv:nucl-ex/9812009].
%%CITATION = NUCL-EX 9812009;%%

\bibitem{Misicu.2001.JPHGB}
  S.~Misicu, M.~Rizea and W.~Greiner,
\newblock
  {\it Emission of electromagnetic radiation in $\alpha$-decay},
\newblock
  Journal of Physics G: Nuclear and Particle Physics {\bf 27},
  993--1003 (2001).
%%CITATION = JPHGB,27,993;%%

\bibitem{Dijk.2003.FBSSE}
  W.~van~Dijk and Y.~Nogami,
\newblock
  {\it Model study of bremsstrahlung in alpha decay},
\newblock
  Few-body systems Supplement {\bf 14}, 229--232 (2003).
%%CITATION = FBSSE,14,229;%%



\bibitem{Serot.1994.NUPHA}
  O.~Serot, N.~Carjan and D.~Strottman,
\newblock
  {\it Transient behaviour in quantum tunneling:
  time-dependent approach to alpha decay},
\newblock
  Nuclear Physics {\bf A 569}, 562--574. (1994).
%%CITATION = NUPHA,569,562;%%

\bibitem{Dijk.1999.PRLTA}
  W.~van~Dijk and Y.~Nogami,
\newblock
  {\it Novel expression for the wave function of a decaying quantum
  systems},
\newblock
  Physical Review Letters {\bf 83}, 2867--2871 (October, 1999).
%%CITATION = PRLTA,83,2867;%%

\bibitem{Dijk.2002.PHRVA}
  W.~van~Dijk and Y.~Nogami,
\newblock
  {\it Analytical approach to the wave function of a decaying quantum
  system},
\newblock
  Physical Review {\bf C65}, 024608--14p. (February, 2002).
%%CITATION = PHRVA,C65,024608;%%

\bibitem{Ivlev.2004.PHRVA}
  B.~Ivlev and V.~Gudkov,
\newblock
  {\it New enhanced tunneling in nuclear processes},
\newblock
  Physical Review {\bf C69}, 037602--4p. (March, 2004);
\newblock
  [nucl-th/0307012].
%%CITATION = NUCL-TH 0307012;%%


\bibitem{Flambaum.1999.PRLTA}
  V.~V.~Flambaum and V.~G.~Zelevinsky,
\newblock
  {\it Quantum Munchhausen effect in tunneling},
\newblock
  Physical Review Letters {\bf 83}, 3108--3111 (1999);
\newblock
  [nucl-th/9812076].
%%CITATION = NUCL-TH 9812076;%%



\bibitem{Maydanyuk.2003.PTP}
  S.~P.~Maydanyuk and V.~S.~Olkhovsky,
\newblock
  {\it Does sub-barrier bremsstrahlung in $\alpha$-decay of
  $^{210}\mbox{Po}$ exist?}
\newblock
  Progress of Theoretical Physics {\bf 109} (2), 203--211 (February, 2003);
\newblock
  arXiv:nucl-th/0404090.
%%CITATION = NUCL-TH 0404090;%%

\bibitem{Maydanyuk.nucl-th.0404013}
  S.~P.~Maydanyuk and S.~V.~Belchikov,
\newblock
  {\it Bremsstrahlung in alpha-decay: angular analysis of spectra}
\newblock
  (talk on the II Conference on High Energy Physics,
  Nuclear Physics and Accelerator Physics,
  March 1-5, 2004, Kharkov, Ukraine),
\newblock
  {\it Problems of atomic science and technology. 
  Series: Nuclear Physics Investigations}
  (44) {\bf 5}, 19-21 (2004); 
\newblock
  arXiv:nucl-th/0404013.
%%CITATION = NUCL-TH 0404013;%%

% \bibitem{Eisenberg.1973}
%   И.~А.~Айзенберг, В.~Грайнер,
% \newblock
%   {\it Механизмы возбуждения ядра},
% \newblock
%   т.~2 (Атомиздат, Москва, 1973), 347~с..

\bibitem{Eisenberg.1973}
  J.~M.~Eisenberg and W.~Greiner,
\newblock
  {\em Mechanisms of nuclear excitations 
  (Mehanizmi vozbuzhdenia yadra)},
\newblock
  Vol.~2 ({Atomizdat}, {Moskva}, 1973), p.~347 --- [in Russian].

\bibitem{Landau.1989}
  L.~D.~Landau and E.~M.~Lifshitz
\newblock
  {\em Quantum mechanics, course of Theoretical Physics},
\newblock
  Vol.~3 ({Nauka}, {Mockva}, 1989), p.~768 ---
  [in Russian; eng. variant: Oxford, Uk, Pergamon, 1982].

\bibitem{Denisov.2005.PHRVA}
  V.~Yu.~Denisov and H.~Ikezoe,
\newblock
  {\it Alpha-nucleus potential for alpha-decay and sub-barrier fusion},
\newblock
  {\it Physical Review} {\bf C72}, 064613 (2005);
\newblock
  [arXiv:nucl-th/0510082].
%%CITATION = PHRVA,C72,064613;%%

\bibitem{Muntyan.2003}
  I.~Muntyan,
\newblock
  {\it Deformed superheavy nuclei},
\newblock
  Ph.D. dissertation
  (Supervisor: Prof.~A.~Sobiczewski, Warsaw, 2003), p.~64.

\bibitem{Amusia.1988.PRPLC}
  M.~Yu.~Amusia,
\newblock
  {\it ``Atomic'' Bremsstrahlung},
\newblock
  Physics Reports {\bf 162}, 249--335 (May, 1988).
%%CITATION = PRPLC,162,249;%%

\bibitem{Amusia.1990}
  M.~Yu.~Amusia,
\newblock
  {\it Tormoznoe izluchenie}
\newblock
  ({Energoatomizdat}, {Moskva}, 1990), p.~208 --- [in Russian].

\bibitem{Amusia.2005}
  M.~Yu.~Amusia,
\newblock
  {\it ``Atomic Bremsstrahlung'': retrospectives, current status and
  perspectives},
\newblock
  [arXiv:physics/0512183].

% \bibitem{Berestetsky.1980}
%   В.~Б.~Берестецкий, Е.~М.~Лифшиц, Л.~П.~Питаевский
% \newblock
%   {\it Квантовая электродинамика},
%   Серия: ``Теоретическая физика'', том 4
% \newblock
%   (Наука, Москва, 1980), 704~с..

% \bibitem{Berestetsky.1980}
%   V.~B.~Berestetsky, E.~M.~Lifshitz and L.~P.~Pitaevsky,
% \newblock
%   {\it Quantum electrodynamics (Kvantovaya elektrodinamika)},
% \newblock
%   Vol.~4 (Nauka, Mockva, 1989), p.~704 --- [in Russian].

\bibitem{Baye.1985.NUPHA}
  D.~Baye and P.~Descouvemont,
\newblock
  {\it Microscopic description of nucleus-nucleus bremsstrahlung},
\newblock
  Nuclear Physics {\bf A 443} (2), 302--320 (1985).
%%CITATION = NUPHA,443,302;%%

\bibitem{Descouvemont.1986.PHLTA}
  P.~Descouvemont,
\newblock
  {\it Microscopic investigation of $\alpha + {}^{16}\mbox{O}$},
\newblock
  Physics Letters {\bf B181} (3--4), 199--202 (December, 1986).
%%CITATION = PHLTA,B181,199;%%

\bibitem{Liu.1990.PHRVA.C41}
  Q.~K.~K.~Liu, Y.~C.~Tang and H.~Kanada
\newblock
  {\it Microscopic calculation of bremsstrahlung emission in
  $^{3}\mbox{He} + \alpha$ collisions},
\newblock
  Physical Review {\bf C41} (4), 1401--1416 (1990).
%%CITATION = PHRVA,C41,1401;%%

\bibitem{Liu.1990.PHRVA.C42}
  Q.~K.~K.~Liu, Y.~C.~Tang and H.~Kanada
\newblock
  {\it Microscopic study of $p + \alpha$ bremsstrahlung},
\newblock
  Physical Review {\bf C42} (5), 1895--1898 (1990).
%%CITATION = PHRVA,C42,1898;%%

\bibitem{Baye.1992.NUPHA}
  D.~Baye, P.~Descouvemont and M.~Kruglanski
\newblock
  {\it Probing scattering wave functions with nucleus-nucleus
  bremsstrahlung},
\newblock
  Nuclear Physics {\bf A 550} (2), 250--262 (December, 1992).
%%CITATION = NUPHA,550,250;%%


\bibitem{Biro.1987.NUPHA}
  T.~S.~Biro, K.~Niita, A.~L.~De~Paoli, W.~Bauer, W.~Cassing and
  U.~Mose,
\newblock
  {\it Microscopic theory of photon production in proton-nucleus and
  nucleus-nucleus collisions},
\newblock
  Nuclear Physics {\bf A 475} (3), 579--597 (December, 1987).
%%CITATION = NUPHA,475,579;%%

\bibitem{Pluyko.1987.FECAA}
  V.~A.~Pluyko and V.~A.~Poyarkov,
\newblock
  {\it Bremsstrahlung in reactions induced by protons},
\newblock
  Physics of Elementary Particles and Atomic Nuclei
  {\bf 18} (2), 374--418 (1987).
%%CITATION = FECAA,18,374;%%

\end{thebibliography}

%%%%%%%%%%%%%%%%%%%%%%%%%%%%%%%%%%%%%%%%%%%%%%%%%%%%%%%%%%%%%%%%%%%%%%%%%%%%%%
%
%       С помощью bib-файла создается bbl-файл-оригинал, который
%       помещается в каталоге c:\EmTeX\Temp (для компьютера в
%       институте ИЯИ).
%
%       Можно вместо bbl-файла-оригинала использовать другой
%       отредактированный bbl-файл, который следует поместить
%       в каталог c:\EmTeX\Temp.
%
%       формат ссылки: <Фамилия автора, год выпуска, CODEN журнала>.
%
%       CODEN:
%               Few body systems Supplement             = FBSSE
%               Nuclear Physics                         = NUPHA
%               Physics Letters                         = PHLTA
%               Physical Review                         = PHRVA
%               PHYS REV LETT                           = PRLTA
%               Soviet Journal of Nuclear Physics
%               (Ядерная физика)                        = SJNCA
%               YADERNAYA FIZ                           = YAFIA
%               Physics of Elementary Particles and
%                 Atomic Nuclei                         = FECAA
%
%       http://www.slac.stanford.edu/spires/coden/
%
%%%%%%%%%%%%%%%%%%%%%%%%%%%%%%%%%%%%%%%%%%%%%%%%%%%%%%%%%%%%%%%%%%%%%%%%%%%%%%
% ***************************************************************************

% ***************************************************************************
\end{document}